%% file: dsfv.tex
\documentclass[11pt,a4paper]{article}

\usepackage[utf8x]{inputenc}
\usepackage{graphicx,a4wide,cite,hyperref}
\usepackage{amsmath,amssymb}

\input{macros}

\title{Quantum electrodynamics in finite volume and nonrelativistic effective field theories}

\author{\begin{minipage}{\textwidth}\begin{center}
Z.\ Fodor$^{1,2,3}$,
C.\ Hoelbling$^{1}$, S.\ D.\ Katz$^{3,4}$, L.\ Lellouch$^{5}$,
A.\ Portelli$^{6}$, K.\ K.\ Szabo$^{1,2}$, B.\ C.\ Toth$^{1}$\\[0.4cm]
{\it $^{1}$\ Department of Physics, University of Wuppertal, D-42119 Wuppertal, Germany\\
$^{2}$\ J\"ulich Supercomputing Centre, Forschungszentrum J\"ulich, D-52428 J\"ulich, Germany\\
$^{3}$\ Institute for Theoretical Physics, E\"otv\"os University, H-1117 Budapest, Hungary\\
$^{4}$\ MTA-ELTE Lend\"ulet Lattice Gauge Theory Research Group, H-1117 Budapest, Hungary\\
$^{5}$ CNRS, Aix Marseille U., U. de Toulon, CPT, UMR 7332, F-13288, Marseille, France\\
$^{6}$\ School of Physics \& Astronomy, University of Southampton, SO17 1BJ, UK
}
\end{center}
\end{minipage}
}

\date{}

\begin{document}

\maketitle
 
\begin{abstract} 

Electromagnetic effects are increasingly being accounted for in
lattice quantum chromodynamics computations. Because of their
long-range nature, they lead to large finite-size effects over which
it is important to gain analytical control. Nonrelativistic effective
field theories provide an efficient tool to describe these
effects. Here we argue that some care has to be taken when applying
these methods to quantum electrodynamics in a finite volume.

\end{abstract}

\vspace{1cm}

State-of-the-art lattice quantum chromodynamics (QCD) computations
have reached such a high level of accuracy (see
e.g.\ \cite{Aoki:2013ldr} and references therein) that small
electromagnetic corrections, and other isospin breaking effects, are
becoming important. These effects will have to be accounted for more
and more systematically if the results of lattice calculations are to
continue to be used to test the standard model and to search for new
physics in increasingly precise experiments. Moreover, electromagnetic
effects in hadrons are important in themselves. For instance, they are
critical for understanding Big Bang nucleosynthesis as well as many
properties of atomic nuclei or for determining the up and down quark
masses. As a result, increasing attention is being focussed on
including quantum electrodynamics (QED) corrections in lattice QCD
calculations. A number of results concerning hadron and quark
  masses have been obtained, in the electroquenched approximation
  \cite{Duncan:1996xy,Blum:2007cy,Blum:2010ym,Borsanyi:2013lga,deDivitiis:2013xla,Drury:2013sfa,Basak:2014vca}
  and in full QCD+QED
  \cite{Aoki:2012st,Ishikawa:2012ix,Horsley:2013qka,Borsanyi:2014jba}. In
  addition, a method for including QED corrections in the lattice
  calculation of hadronic matrix elements has been proposed in
  \cite{Carrasco:2015xwa}.

A very important issue in such calculations is finite-volume
effects.~\footnote{Throughout this paper we will be concerned with
  situations in which the linear dimensions of the lattice are much
  larger than the Compton wavelengths and the internal-structure
  length-scales of the particles under consideration.}  Indeed, the
vanishing mass of the photon implies that the leading finite-volume
corrections are proportional to inverse powers of the spatial
dimension, $L$, of the lattice, instead of being exponentially
suppressed, as they are in pure QCD calculations. These power
corrections represent a large fraction--tens of percent--of the
computed electromagnetic effects for typical lattice sizes $L\sim
3\,\fm$, as shown in
\cite{Borsanyi:2013lga,Borsanyi:2014jba,Basak:2014vca}, where
controlled infinite-volume extrapolations were performed. These
corrections must be appropriately subtracted to obtain accurate
results. Thus, it is important to know their precise analytical form
and to constrain them as much as possible.~\footnote{The two leading
  finite-volume corrections, proportional to $1/L$ and $1/L^2$, are
  given in terms only of the particle's charge and infinite-volume
  mass \cite{Davoudi:2014qua,Borsanyi:2014jba}. This important feature
  allows precise infinite-volume extrapolations of the QED
  contributions to particle masses in lattice calculations. In
  particular, the finite-volume corrections do not depend on spin nor
  on particle structure \cite{Davoudi:2014qua,Borsanyi:2014jba}. As
  shown in \cite{Borsanyi:2014jba}, this {\em universality} follows,
  under very general hypotheses, from QED Ward-Takahashi identities
  and the work of L\"uscher on the analyticity properties of
  propagators and vertex functions \cite{Luscher:1985dn}.}

An elegant and efficient method to determine the functional form of
these effects was proposed and worked out in
\cite{Davoudi:2014qua}. It is based on nonrelativistic effective field
theories (NREFTs) and allows to compute these corrections in a
systematic expansion in powers of $1/L$ for any spin-$0$ or $1/2$
massive particles with (or without) internal structure, such as
hadrons or nuclei. In this approach, the corrections are determined by
the mass and by well-defined electromagnetic properties of the particle,
such as its charge, charge radius, magnetic moment, etc. However, as
pointed out in \cite{Borsanyi:2014jba} and explained in more detail
below, the results obtained in \cite{Davoudi:2014qua} do not fully
agree with those of point-particle calculations performed in
\cite{Borsanyi:2014jba}. Since many checks of the point-particle
calculation were performed in \cite{Borsanyi:2014jba} and both results
cannot be correct, we have been led to investigate the possible source
of the discrepancy. As explained below, this investigation has led us
to uncover a subtlety in the application of NREFTs to finite-volume
QED, which goes beyond the calculation discussed in the present
paper.~\footnote{It is important to note that this subtlety does not
  affect the two leading, {\em universal}, finite-volume corrections
  mentioned above, as it only enters at order $1/L^3$, as shown
  below.}

To expose the problem and its solution, we focus on the calculation of
the finite-volume corrections to the pole mass of spin-$1/2$ particles
at $O(\alpha)$. As in \cite{Davoudi:2014qua,Borsanyi:2014jba}, we work
with periodic boundary conditions in the $\qedL$ formulation of QED in
a finite volume. In this formulation, momentum modes of the photon
field with $\vec{k}=\vec{0}$, for all values of $k_0$, are eliminated
from the theory. This approach was first proposed in
\cite{Hayakawa:2008an} and, as shown in \cite{Borsanyi:2014jba}, has
many theoretical and practical advantages. Here we work with an
infinite time direction, as it simplifies analytical calculations and
is equivalent to the finite time extent, $T$, formulation with
periodic boundary conditions, up to corrections which are smaller than
any inverse power in $T$ \cite{Borsanyi:2014jba}.

Finite-volume, electromagnetic corrections to the mass of a particle
can be obtained from the difference of its on-shell, electromagnetic
self-energy at rest, in finite and infinite volumes. Physically, the
corrections in inverse powers of $L$ arise mainly from the particle
under consideration exchanging a photon with itself around the
periodic three-volume. In \cite{Davoudi:2014qua}, Davoudi and Savage
(DS) compute this effect for a generic spin-$1/2$ particle using NREFT
to N$^3$LO, i.e. to order $1/L^4$. The use of NREFT is entirely
justified here, because we are after an asymptotic expansion of
particle properties in inverse powers of $L$. NREFT is a low-momentum
effective theory of QED and in $\qedL$, it
provides expansions of low-momentum particle properties in powers of
the infrared momentum cutoff, $2\pi/L$.

\begin{figure}
 \centering 
\includegraphics[width=\textwidth]{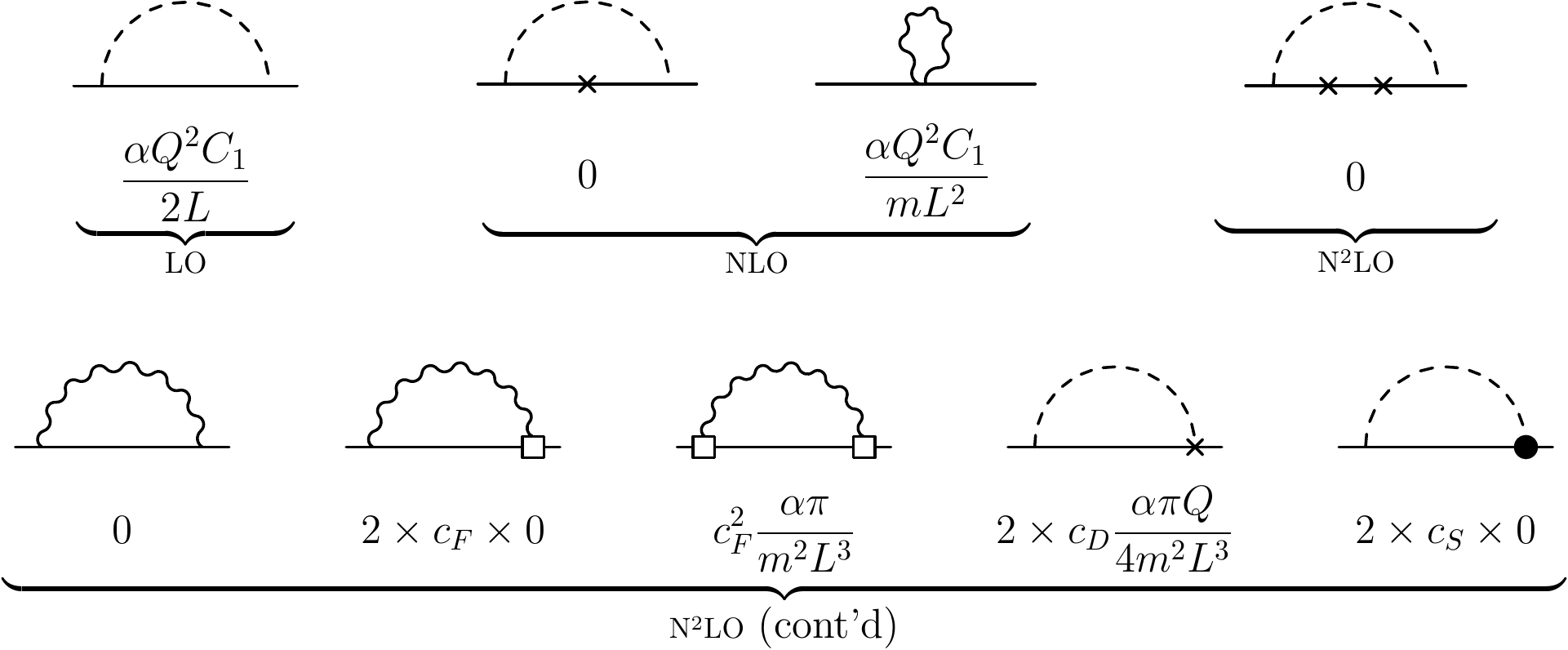}
\caption{\label{fig:NREFT_SE_N2LO}\em NREFT self-energy diagrams which
  contribute to finite-volume effects on the mass of a spin-1/2
  particle, up to order $1/L^3$. Their contributions were taken into
  account in \protect\cite{Davoudi:2014qua}. The solid line
  corresponds to the fermion propagator; the dashed line to the
  temporal photon propagator; the wavy line to the transverse photon
  propagator; the cross to a kinetic term insertion; the simple
  vertices with these lines, to the fermion-temporal-photon
  interaction; to the fermion-transverse-photon interaction; to the
  fermion-to-two-transverse-photons interaction; the square vertex to
  the Fermi interaction; the crossed vertex to the Darwin interaction;
  the dot vertex to the spin-orbit interaction. Under each diagram we
  indicate its contribution to the finite-volume correction on the
  fermion mass. Calculations are performed in Coulomb gauge. An
  explicit factor of 2 appears in the single Fermi, Darwin and
  spin-orbit vertex terms. This indicates that to each of the three
  diagrams shown corresponds one in which the vertices are
  switched. These new diagrams give the same contributions as the
  original diagrams. The sum of all the contributions shown in the
  figure yields the result in \protect\eq{eq:dsfv}.}
\end{figure}

To understand the NREFT approach, we have repeated their
calculation. We stop at N$^2$LO, i.e. up to and including terms
proportional to $1/L^3$, because this is the order at which the
subtlety, described below, first arises. We use the same NREFT
lagrangian as they do,
namely~\cite{Caswell:1985ui,Thacker:1990bm,Labelle:1992hd,Manohar:1997qy,Luke:1997ys,Chen:1999tn,Hill:2012rh,Lee:2014iha}:
\bea
\label{eq:Lpsi}
\cL_\psi & = & \psi^\dagger\left[i D_0 + \frac{|\vec{D}|^2}{2m}+c_F
  \frac{e}{2m}\vec{\sigma}.\vec{B}+c_D\frac{e}{8m^2}\vec{\nabla}\cdot\vec{E}\right.\nn\\
  & & \left. + ic_S\frac{e}{8m^2}
  \vec{\sigma}\cdot(\vec{D}\times\vec{E}-\vec{E}\times\vec{D})+O(\vec{p}^4)\right]\psi \ ,
\eea
where $\psi$ is a two-component spinor which annihilates a particle,
$D_\mu=\partial_\mu+ieQ A_\mu$, $c_F=Q+\kappa+O(\alpha)$,
$c_D=Q+4m^2\langle r^2\rangle/3+O(\alpha)$, $eQ$ is the charge of the
particle, $m$ its infinite-volume mass, $\langle r^2\rangle$ its
mean-squared charge radius and $\kappa$ its anomalous magnetic
moment. The diagrams which contribute up to N$^2$LO are shown in
\fig{fig:NREFT_SE_N2LO}. They yield the following finite-volume
corrections:
\be \Delta_{\mbox{\tiny DS}} m(L)\equiv m(L)-m \underset{L\to+\infty}{\sim}
\alpha Q^2\frac{C_1}{2L}\left[1+\frac{2}{mL}\right]
+\frac{\pi\alpha}{m^2L^3}\left[c_F^2+\frac12 c_D Q\right]
\ ,
\label{eq:dsfv}
\ee
where $m(L)$ is the value of the particle's mass in $\qedL$ and $C_1=-2.837297(1)$
\cite{Luscher:1986pf,Hasenfratz:1989pk,Luscher:1990ux,Borsanyi:2014jba}
is a known constant. This result fully agrees with DS's.

However, as noted above and in \cite{Borsanyi:2014jba}, if we reduce
the corrections of \eq{eq:dsfv} to the point particle case by setting
$\langle r^2\rangle = 0$ and $\kappa = 0$, we find that it disagrees
with the one that we computed directly in spinor QED. Indeed, in the
reduced corrections, the term proportional to $(\pi\alpha Q^2/m^2L^3)$
has a coefficient of $3/2$ instead of $3$, as found in
\cite{Borsanyi:2014jba}. Moreover, in \cite{Borsanyi:2014jba} we
performed a precise, dedicated numerical study of finite-volume
effects on the pole mass of a point-like fermion in $\qedL$. The study
was conducted with fixed bare parameters and six lattice sizes ranging
from $L/a=24$ to $L/a=128$, with $a$ the lattice spacing. A fit of the
measured, finite-volume mass, to a polynomial in $a/L$, highly favored
our value for the $1/L^3$ coefficient over the one in
\cite{Davoudi:2014qua}. To explain where this discrepancy comes from,
we must now describe the main features of our
relativistic calculation \cite{Borsanyi:2014jba}.

The computation in spinor QED is a straightforward asymptotic
expansion in large $L$, of the difference between the finite and
infinite-volume, on-shell self-energies of the particle
\cite{Borsanyi:2014jba}. As mentioned above, we consider this
difference at $O(\alpha)$, i.e.\ we consider the usual, one-loop sunset diagram in
which a photon is emitted and re-absorbed by the particle. The
asymptotic expansion is most straightforwardly performed using a
Poisson summation formula. In that approach, the corrections in powers
of $1/L$ result from the nonanalyticities in the integrand/summand,
associated with intermediate states going on shell in the domain of
integration. The obvious singularities that arise in the present case
are the particle and the positive and negative energy photon
poles. The antiparticle pole, which is $2m$ away in energy, is only
expected to contribute terms which fall off exponentially with
$2mL$. This is an illustration of the decoupling of antiparticle modes
that leads to the NREFT for a single, massive particle, used in
\cite{Davoudi:2014qua}.

Upon closer inspection, however, we find that this expectation is
incorrect. Analyzing the different contributions shows that the
antiparticle pole contributes to the contentious term of order
$1/(m^2L^3)$, even though its propagator cannot go on shell for the
given kinematics. This surprising result is due to the fact that the
contribution of the zero modes of the photon are omitted from the loop
sum. At order $1/L^3$, the finite-volume corrections come not only
from singularities of the summand/integrand in the domain of
integration, but also from the explicit subtraction of the photon zero
modes. Since these modes couple particles to antiparticles, the latter
also play a role in the calculation of finite-volume effects.

In the language of NREFT, we reach the same conclusions if we
  explicitly include antiparticle degrees of freedom, in addition to
  the usual contributions of antiparticles to higher-dimension
  particle operators. Thus, to the lagrangian density, $\cL_\psi$,
for particle fields given in \eq{eq:Lpsi} and used in
\cite{Davoudi:2014qua}, we add the one for antiparticle fields,
$\chi$, and include direct couplings of particles to
antiparticles. Then, the lagrangian density becomes:
\be
\label{eq:cLtot}
\cL_\mathrm{N^2LO} = \cL_\psi + \cL_\chi + \cL_{4f} + O(\vec{p}^4)
\ ,\ee
where $\chi$ is a two-component spinor which annihilates
antiparticles. In \eq{eq:cLtot}, $\cL_\chi$ is obtained from
$\cL_\psi$ with the replacements $\psi\to\chi$ and $Q\to -Q$. The
relevant four-fermion interaction, at $O(\alpha)$, originates from
particle-antiparticle annihilation in the triplet channel and is given
by \cite{Labelle:1997uw}
\be
\label{eq:cLpsichi}
\cL_{4f} = 
d_V\frac{\alpha}{m^2}(\psi^\dagger\vec{\sigma}\sigma_2\chi^*)\cdot(\chi^T\sigma_2\vec{\sigma}\psi)
+O(\alpha^2,\vec{p^4})
\ ,\ee
where the $\sigma_i$, $i=1,2,3$, are the Pauli matrices. For point
particles, $d_V=-\pi Q^2+O(\alpha)$
\cite{Caswell:1985ui,Kinoshita:1990ai}.

\begin{figure}
  \centering
\includegraphics[width=0.25\textwidth]{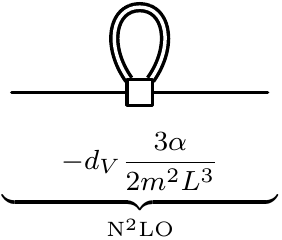}
  \caption{\label{fig:NREFT_SE_4f}\em Antiparticle contribution to the
    self-energy of a spin-1/2 particle which arises from the
    four-fermion lagrangian of \eq{eq:cLpsichi}. The double line
    corresponds to the antiparticle propagator. The vertex corresponds
    to the vector four-fermion coupling. Under the diagram we indicate its
    contribution to the finite-volume corrections on the fermion
    mass. The diagram contributes at order $1/L^3$ and provides the
    term which is missing in the calculation of
    \protect\cite{Davoudi:2014qua}.}
\end{figure}

From this four-fermion lagrangian, it is straightforward to compute
the contribution of the antiparticle to the finite-volume effects in
the particle's mass. It is given by the self-energy diagram 
of \fig{fig:NREFT_SE_4f}. We find this
contribution to be
\be
\label{eq:DmL_psichi}
\Delta_{4f} m(L) =
 d_V\frac{3\alpha}{2m^2L^3}\;\widehat{\sum_{\vec{q}\ne
    \vec{0}}}\; 1 \ ,  \ee
where $\vec{q}$ is the momentum of the antiparticle in the loop and the sum,
$\widehat{\sum}_{\vec{q}\ne
    \vec{0}}$, represents the difference between the sum over the
finite-volume modes and the infinite-volume integral, i.e.
\be
\label{eq:FVsumint}
\frac1{L^3}\;\widehat{\sum_{\vec{q}\ne \vec{0}}}
\equiv
\frac1{L^3}\sum_{\vec{q}\ne \vec{0}}-
\int \frac{d^3q}{(2\pi)^3}\ .
\ee
We now argue that the $\vec{q}=\vec{0}$ modes must be eliminated from
the finite-volume sum, even though it is, initially, only photon zero
modes which are removed. This is where the subtlety enters in
NREFT. If the $\vec{q}=\vec{0}$ antiparticle modes were present, as
one might guess, then the finite-volume correction of
\eq{eq:DmL_psichi} would vanish and antiparticle degrees of freedom
would not contribute. However, one must remember
where internal particle or antiparticle lines in NREFT come from. In
relativistic QED, an internal particle or antiparticle line in a
diagram, such as the self-energy diagram under consideration, is
produced at a vertex with a photon. But in $\qedL$ that photon cannot
have a vanishing three-momentum. Therefore in such diagrams, where all
external particle lines have vanishing three-momenta, the internal
antiparticle lines cannot have vanishing three-momenta. This justifies
the omission of the $\vec{q}=\vec{0}$ antiparticle modes in the
contribution of \eq{eq:DmL_psichi}.

Now $\widehat{\sum}_{\vec{q}\ne \vec{0}}\; 1=-1$, so the full NREFT
expression for the finite-volume correction to the mass $m$ of a
spin-$1/2$ particle of charge $Q$ is, to $O(1/L^3)$ in the
presence of electromagnetism,
\be \Delta m(L) \underset{L\to+\infty}{\sim}
\Delta_{\mbox{\tiny DS}} m(L) - d_V\frac{3\alpha}{2m^2L^3}
\ .
\label{eq:correctfv}
\ee
Using the value of $d_V$ for a point particle given after
\eq{eq:cLpsichi}, we find that this additional correction adds to the
ones in $\Delta_{\mbox{\tiny DS}} m(L)$ exactly the $(3\pi\alpha
Q^2/2m^2L^3)$ term which is missing to reproduce the relativistic
point-particle result of \cite{Borsanyi:2014jba}.

It is worth noting that the result of \eq{eq:correctfv} can also be
obtained by only reinstating the $\vec{q}=\vec{0}$ modes of the
antiparticle field in the finite-volume NREFT. Then one uses the
corresponding terms in the lagrangian to subtract these modes'
contribution from the finite-volume self-energy. In that approach, the
$\widehat{\sum}_{\vec{q}\ne \vec{0}}\; 1$ that appears in
\eq{eq:DmL_psichi} would directly be replaced by $-1$. One need
consider neither the spatial modes of the antiparticle field in finite
volume, nor any antiparticle modes in infinite volume.

To conclude, we have considered the calculation of finite-volume
corrections to the pole mass of a charged spin-$1/2$ particle, in the
presence of electromagnetism. We have explained how the NREFT
calculation of \cite{Davoudi:2014qua} can be reconciled with the
relativistic QED result of \cite{Borsanyi:2014jba}. In the process, we
have shown that there are subtleties associated with applying NREFTs
to the calculation of finite-volume effects in QED. In particular, we
have argued that antiparticle degrees of freedom must be dealt with
carefully, because photon zero modes are treated differently in finite
and infinite volumes. Indeed, those modes can couple antiparticles to
particles. Therefore, NREFT calculations of particle properties must
account for the contribution of antiparticles to the removal of the
photon zero modes in finite volume. However, once this contribution
is suitably accounted for, the NREFT approach constitutes an elegant
and efficient formalism to calculate finite-volume corrections.

\subsection*{Acknowledgments}

This work was supported in part by the OCEVU Labex
(ANR-11-LABX-0060) and the A$^\star$MIDEX project
(ANR-11-IDEX-0001-02), funded by the ``Investissements d'Avenir''
French government program and managed by the ANR, by DFG grants FO
502/2 and SFB-TR 55 and by STFC grants ST/J000396/1 and ST/L000296/1.

\bibliographystyle{JHEP_notitle}
\bibliography{/Users/lellouch/Bibliography/all_papers}

\end{document}

%% file: macros.tex
\def\be{\begin{equation}}
\def\ee{\end{equation}}
\def\bea{\begin{eqnarray}}
\def\eea{\end{eqnarray}}

\def\reff#1{\ref{#1}}

\def\eq#1{Eq.~(\reff{#1})}

\def\fig#1{Fig.~\reff{#1}}

\def\nn{\nonumber}

\newcommand{\lsim}{ {\
\lower-1.2pt\vbox{\hbox{\rlap{$<$}\lower5pt\vbox{\hbox{$\sim$}}}}\ } }
\newcommand{\gsim}{ {\
\lower-1.2pt\vbox{\hbox{\rlap{$>$}\lower5pt\vbox{\hbox{$\sim$}}}}\ } }

\def\er#1#2{\relax\ifmmode{}^{+#1}_{-#2}\else$^{+#1}_{-#2}$\fi}
\def\erparen#1#2{\relax\ifmmode{}(^{#1}_{#2})\else$(^{#1}_{#2})$\fi}

\def~{\ifmmode\phantom{0}\else\penalty10000\ \fi}

\def\fm{\mathrm{fm}}
\def\ev{\mathrm{e\kern-0.1em V}}
\def\kev{\mathrm{ke\kern-0.1em V}}
\def\mev{\mathrm{Me\kern-0.1em V}}
\def\gev{\mathrm{Ge\kern-0.1em V}}
\def\tev{\mathrm{Te\kern-0.1em V}}

\def\n#1e#2n{{#1}\times 10^{#2}}

\def\nn{\nonumber}

\def\cL{\mathcal{L}}

\def\ods2{\mathcal{O}_{\Delta S=2}}
\def\zds2{Z_{\Delta S=2}}

\makeatletter
\def\slash#1{{\mathpalette\c@ncel{#1}}} % TeXbook, bottom of p360
\def\big#1{{\hbox{$\left#1\vbox to1.012\ht\strutbox{}\right.\n@space$}}}
\def\Big#1{{\hbox{$\left#1\vbox to1.369\ht\strutbox{}\right.\n@space$}}}
\def\bigg#1{{\hbox{$\left#1\vbox to1.726\ht\strutbox{}\right.\n@space$}}}
\def\Bigg#1{{\hbox{$\left#1\vbox
to2.083\ht\strutbox{}\right.\n@space$}}}
\makeatother

\makeatletter
\def\old@comma{,}
\catcode`\,=13
\def,{%
  \ifmmode%
    \old@comma\discretionary{}{}{}%
  \else%
    \old@comma%
  \fi%
}
\makeatother

\makeatletter
\def\old@semicolon{;}
\catcode`\;=13
\def;{%
  \ifmmode%
    \old@semicolon\discretionary{}{}{}%
  \else%
    \old@semicolon%
  \fi%
}
\makeatother

\def\qedL{\mathrm{QED}_L}

\usepackage{xcolor}